\documentstyle[prl,aps,multicol]{revtex}
\begin{document}

\draft

\title
{Spontaneous spin current near the interface between
unconventional superconductors and ferromagnets 
}

\author{Kazuhiro Kuboki and Hidenori Takahashi}

\address{Department of Physics, Kobe University, Kobe 657-8501, Japan}

\date{\today}
%\date{  }

\maketitle 

\begin{abstract} 
We study theoretically the proximity effect between ferromagnets (F) and 
superconductors (S) with broken time-reversal symmetry (${\cal T}$).   
A chiral $(p_x\pm ip_y)$-wave,  and a $d_{x^2-y^2}$-wave 
superconductor,  the latter of which can form  
${\cal T}$-breaking surface state, i.e., $(d_{x^2-y^2} \pm is)$-state, 
are considered for the S side. 
The spatial variations of the superconducting order parameters and the 
magnetization are determined by solving the Bogoliubov de Gennes equation. 
In the case of a chiral $(p_x\pm ip_y)$-wave superconductor, 
a spontaneous spin current flows along the interface, but not 
in the case of a $d_{x^2-y^2}$-wave superconductor. 
For F/S$(p_x \pm ip_y)$/F trilayer system, 
total spin current can be finite while total charge current vanishes,  
if the magnetization of two F layers are antiparallel.

\end{abstract}


%%%%%%%%%%%
\begin{multicols}{2}

\narrowtext

\section{Introduction}

Recently the proximity effect of unconventional superconductors 
has been a subject of intensive study
\cite{layer,Tana,Demler,add6,KK1,Zhu,Amin,Amin2}.
This is because the interface properties of these superconductors 
can be quite different from those of conventional ($s$-wave) 
ones due to the nontrivial angular structure of pair wave 
functions, so that their study is of particular interest.  
The proximity effect between unconventional superconductors and 
magnetic materials is an important problem, since there are competition 
and the possible coexistence of two kinds of orders.
The unconventional superconductivity is usually induced by the 
magnetic interaction, and the existence of nodes 
in the gap is favorable for the coexistence compared with the case 
of conventional superconductors. 
If the magnetism and superconductivity coexist 
near the interface, the electronic properties of the surface state 
become quite unusual. In particular the spin-triplet superconducting 
order parameter (SCOP) 
can be induced near the surface of spin-singlet superconductors\cite{KK1}.
 (Similar effects for conventional 
$s$-wave superconductors and ferromagnets have been studied in 
\cite{Berg2,Volkov,Berg}.) 

In this paper we study the proximity effect between ferromagnets 
and unconventional superconductors with $d_{x^2-y^2}$-wave and 
$(p_x \pm ip_y)$-wave symmetries. 
The former state is realized in high-$T_c$ cuprates\cite{Scal}, 
while the latter is a candidate of the superconducting (SC) state 
of Sr$_2$RuO$_4$\cite{RS}.
It is known that near the [110] surface of the $d_{x^2-y^2}$-wave 
superconductor the system may break 
time-reversal symmetry (${\cal T}$) by introducing second component of 
SCOP\cite{Sig,SBL,KK2,Matsu,F,Amin,Amin2},  
and that a spontaneous current and fractional vortices may occur at 
the surface if ${\cal T}$ is 
broken\cite{Vol,SigR,SigUe,Sig,SBL,KK2,Matsu,F,Amin,Amin2}.
The $(p_x \pm ip_y)$-wave SC states also break ${\cal T}$ and 
the spontaneous current arises at the edge of the system\cite{matsu2}.
When a ferromagnet is attached to these unconventional superconductors, 
the magnetization may be induced in the latter due to the proximity 
effect\cite{KK1}.
Then it may be expected that a spontaneous spin current can appear 
along the interface between a ${\cal T}$-breaking superconductor and 
a ferromagnet, because there is an imbalance of the densities of spin-up 
and spin-down electrons. 
We will show that it is actually possible for the superconductor 
with $(p_x \pm ip_y)$-wave symmetry, but not in the case of  $d_{x^2-y^2}$-wave
symmetry\cite{LT23}.
(Spin currents in the case of interfaces between 
$s$-wave superconductors and ferromagnets have been found 
in \cite{Kraw}, though their treatment of the magnetization is not 
fully self-consistent.)

\section{Model and BdG Equations}

The system we consider is a two-dimensional 
ferromagnet(F)/superconductor(S) bilayer and 
F/S/F trilayer systems, and 
we treat [100] and [110] interfaces as schematically depicted in Fig.1. 
The directions perpendicular (parallel) to the interfaces are denoted 
as $x$ ($y$) and $x'$ ($y'$), for [100] and [110] interfaces, respectively. 
The axes $x'$ and $y'$ are 45$^\circ$ rotated from the  crystal axes directions 
$x$ and $y$.  We assume that the system 
is uniform along the direction parallel to the interface. 
In order to describe the magnetism and superconductivity,  
the tight-binding model on a square lattice with on-site repulsive and 
nearest-neighbor attractive interactions is treated within the mean-field 
(MF)  approximation\cite{KK1,Zhu}, 
and we consider only the case of zero temperature ($T = 0$). 
The Hamiltonians for the two layers are given by 
\begin{equation} 
\begin{array}{rl}
  H_L & =  - \displaystyle t_L \sum_{<i,j> \sigma} 
    ( c_{i,\sigma}^{\dagger} 
      c_{j,\sigma} + h.c.)  
     +  U_L \sum_i n_{i\uparrow}n_{i\downarrow}  \\ 
     + & \displaystyle 
     V_L \sum_{<i,j>} \big[n_{i\uparrow}n_{j\downarrow}  
     + n_{i\downarrow}n_{j\uparrow} \big], 
     \ \ (L = F, S) 
\end{array}     
\end{equation}
where $\langle i,j \rangle $ denotes the nearest-neighbor bonds, 
$n_{i\sigma} \equiv c_{i\sigma}^\dagger c_{i\sigma}$,
and $c_{i\sigma}$ is the annihilation operator of electrons at a site $i$ 
with spin $\sigma$.
Parameters $t_L$, $U_L$ and $V_L$ represent the 
transfer integral, the on-site interaction and the nearest-neighbor
interaction, respectively, for the $L (=F, S)$ side. 
The transmission of electrons at the interface is described by the 
following tight-binding Hamiltonian
\begin{equation}
    H_T =  - \displaystyle t_T \sum_{<l,m>\sigma} 
    ( c_{l,\sigma}^{\dagger} 
      c_{m,\sigma} + h.c.)     
\end{equation}
where $l$ ($m$) denotes the surface sites of $F$ ($S$) layer, 
and then the total Hamiltonian of the system is $H = H_F + H_S + H_T 
- \mu \sum_{i\sigma} c_{i\sigma}^\dagger c_{i\sigma}$ with 
$\mu$ being the chemical potential.  
The interaction terms are decoupled within the MF approximation:
\begin{equation}
\begin{array} {rl}
U n_{i\uparrow}n_{i\downarrow} \to &
U\langle n_{i\uparrow} \rangle n_{i\downarrow} 
+ U\langle n_{i\downarrow} \rangle n_{i\uparrow}
-U \langle n_{i\uparrow} \rangle \langle n_{i\downarrow} \rangle, \\
Vn_{i\uparrow}n_{j\downarrow} \to &
V\Delta_{ij}c_{j\downarrow}^\dagger c_{i\uparrow}^\dagger 
+V\Delta_{ij}^{*}c_{i\uparrow} c_{j\downarrow}
-V\vert \Delta_{ij} \vert^2 
\end{array}
\end{equation}
with $\Delta_{ij} \equiv \langle c_{i\uparrow}c_{j\downarrow}\rangle$.
Then $\Delta_{ij}$ and the magnetization 
$m_i =\langle n_{i\uparrow} - n_{i\downarrow} \rangle/2$ 
(perpendicular to the plane) are the OP's to be determined self-consistently. 
The SCOP with each symmetry can be formed by combining 
$\Delta_{ij}$'s: 
$\Delta_d(i) \equiv  (\Delta_{i,i+{\hat x}}^{(S)} + \Delta_{i,i-{\hat x}}^{(S)}
- \Delta_{i,i+{\hat y}}^{(S)} - \Delta_{i,i-{\hat y}}^{(S)})/4$ ($d_{x^2-y^2}$-wave),  
$\Delta_s(i) \equiv  (\Delta_{i,i+{\hat x}}^{(S)} + \Delta_{i,i-{\hat x}}^{(S)} 
+ \Delta_{i,i+{\hat y}}^{(S)} + \Delta_{i,i-{\hat y}}^{(S)})/4$ (extended $s$-wave ),  
$\Delta_{px(y)}(i) \equiv (\Delta_{i,i+{\hat x}({\hat y})}^{(T)} 
- \Delta_{i,i-{\hat x}({\hat y})}^{(T)})/2$ ($p_{x(y)}$-wave), 
where $\Delta_{ij}^{(S)} \equiv (\Delta_{ij} + \Delta_{ji})/2$ and 
$\Delta_{ij}^{(T)} \equiv (\Delta_{ij} - \Delta_{ji})/2$ are the spin-singlet 
and the spin-triplet pairing OP's, respectively. 

We impose the open (periodic) boundary condition for the $x$ and $x'$ 
($y$ and $y'$) directions for [100] and [110] interfaces, respectively, 
and carry out the Fourier transformation along the $y$- and $y'$-directions. 
For the [100] case we define 
$\Delta^{x\pm}_{x_i} = \Delta_{i,i\pm{\hat x}}$ and 
$\Delta^{y\pm}_{x_i} = \Delta_{i,i\pm{\hat y}}$ as the SCOP's independent of $y$. 
Similarly for the [110] case, $\Delta^{\alpha\beta}_{x'_i} = 
\Delta_{i,i +\alpha \frac{{\hat x}'}{2} +\beta \frac{{\hat y}'}{2}}$ 
($\alpha$, $\beta = +$ or $-$) are SCOP's independent of $y'$.
($|{\hat x}| = |{\hat y}| = a$, $|{\hat x}'| = |{\hat y}'| = a'$ with 
$a$ and $a'$ defined in Fig.1.)  
Then the mean-field Hamiltonian is written as  
(hereafter $i$ denotes $x_i$ and $x_i^{'}$ for a [100] and a [110] interface, 
respectively)
\begin{equation}
{\cal H}_{\rm MFA} = \sum_k\sum_i\sum_j \Psi_i^\dagger(k) 
{\hat h}_{ij}(k) \Psi_j(k)
\end{equation} 
with $\displaystyle \Psi_i^\dagger(k) 
= \big(c_{i\uparrow}^\dagger(k), c_{i\downarrow}(-k)\big)$, and 
$k$ is the wave number for the direction parallel to the interface. 
The matrix ${\hat h}_{ij}(k)$ is given as
\begin{equation}
\displaystyle {\hat h}_{ij}(k) = 
\left (\begin{array}{cc}
\xi_{ij\uparrow}(k)  & F_{ij}(k)  \\
F_{ij}^{*}(k) & - \xi_{ij\downarrow}(k)
\end{array}\right ), 
\end{equation}
where 
\begin{equation}\begin{array}{rl}
\displaystyle \xi_{ij\sigma}(k) = & 
(-2t_{ij}\cos ka -\mu + U\langle n_{i,-\sigma}\rangle)\delta_{ij} 
\\ 
- & \displaystyle t_{ij} (\delta_{i,j+a} + \delta_{i,j-a}),   
\\
& \\
\displaystyle F_{ij}(k) = & V_{ij}[\Delta^{x+}_i\delta_{i,j-a} 
+ \Delta^{x-}_i\delta_{i,j+a}
\\ & \\
+ & \displaystyle (\Delta^{y+}_i e^{{\rm i}ka} 
+ \Delta^{y-}_i e^{-{\rm i}ka})\delta_{i,j}]
\end{array}\end{equation}
for the [100] interface, and 
\begin{equation}\begin{array}{rl}
\displaystyle \xi_{ij\sigma}(k) = 
& (-\mu + U\langle n_{i,-\sigma}\rangle)\delta_{ij} 
\\ 
- & \displaystyle 2t_{ij}\cos(\frac{ka'}{2}) (\delta_{i,j+\frac{a'}{2}} 
+ \delta_{i,j-\frac{a'}{2}})  
\\
& \\
\displaystyle F_{ij}(k) = & V_{ij}[
\delta_{i,j-\frac{a'}{2}}(\Delta^{++}_ie^{{\rm i}ka'/2} 
+ \Delta^{+-}_ie^{-{\rm i}ka'/2}) 
\\ & \\
+ & \displaystyle \delta_{i,j+\frac{a'}{2}}
(\Delta^{-+}_ie^{{\rm i}ka'/2} + \Delta^{--}_ie^{-{\rm i}ka'/2})]
\end{array}\end{equation}
for the [110] interface. Here $t_{ij} = t_{F(S)}$, $V_{ij} = V_{F(S)}$ 
if both $i$ and $j$ are on the F (S) side, while $t_{ij} =t_T$, $V_{ij} = 0$ 
if $i$ and $j$ correspond to the interface sites.

We diagonalize the mean-field Hamiltonian by solving the following
Bogoliubov de Gennes (BdG) equation\cite{deG}:
\begin{equation} 
\sum_j {\hat h}_{ij}(k) 
\left (\begin{array}{cc}
u_{jn}(k)   \\
v_{jn}(k)
\end{array}\right )
= E_n(k) 
\left (\begin{array}{cc}
u_{in}(k)   \\
v_{in}(k)
\end{array}\right )  
\end{equation} 
where $E_n(k) $ and $(u_{in}(k), v_{in}(k))$ are the energy eigenvalue and the 
corresponding eigenfunction, respectively, for each $k$. 
The unitary transformation using $(u_{in}(k), v_{in}(k))$ 
diagonalizes the matrix ${\cal H}_{\rm MFA}$, and  conversely 
the OP's $\Delta_{ij}$ and $m_i$ can be written in terms of  
$E_n(k) $ and $(u_{in}(k), v_{in}(k))$:
\begin{equation}\begin{array}{rl}
\Delta_i^{x\pm} = & \displaystyle \frac{1}{N_y} \sum_{k,n} 
u_{i,n}(k)v_{i\pm a,n}^*(k)[1-f(E_n(k))] 
\\
\Delta_i^{y\pm} = & \displaystyle \frac{1}{N_y} \sum_{k,n} 
u_{i,n}(k)v_{i,n}^*(k)[1-f(E_n(k))]e^{\mp {\rm i}ka} 
\\
\Delta_i^{\alpha\beta} = & \displaystyle \frac{1}{N_{y'}} \sum_{k,n} 
u_{i,n}(k)v_{i+\alpha \frac{a'}{2} ,n}^*(k)
%\\ & \displaystyle * 
[1-f(E_n(k))]  e^{-{\rm i}\beta ka'/2}
\\
\langle n_{i\uparrow} \rangle = & \displaystyle 
\frac{1}{N_y} \sum_{k,n} |u_{i,n}(k)|^2 f(E_n(k)) \\ 
\langle n_{i\downarrow} \rangle = & \displaystyle 
\frac{1}{N_y} \sum_{k,n} |v_{i,n}(k)|^2 [1-f(E_n(k))] 
\end{array}\end{equation}
where $f(E)$ is the Fermi distribution function.
(For the [110] case, $N_y$ in the last two expressions should be 
replaced by $N_{y'}$.)  
These constitute the self-consistency equations which  
will be solved numerically in the following. 

The procedure of the self-consistent numerical calculation is the following. 
We substitute an initial set of OP's in the matrix elements of ${\hat h}_{ij}(k)$,  
and solve BdG equation (Eq.(8)) to get eigenvalues and eigenfunctions. 
Then we recalculate the OP's, and the iteration is performed until the values of 
OP's are converged. We have used various sets of initial OP's 
for the same values of the parameters of the model. 
If several different solutions are obtained,  
we adopt the solution with the lowest energy as the true one.

In the uniform case with $U > 0$ and $V = 0$ (i.e., repulsive Hubbard model), 
the ground state phase diagram within the MF approximation was examined 
by Hirsch\cite{Hirsch}. We use this result to choose the parameters to realize 
the ferromagnetic state in the F layer.  The phase diagram in the case 
of $U = 0$ and $V < 0$ was also studied, and the $d_{x^2-y^2}$-, 
extended $s$- and $(p_x \pm ip_y)$-wave SC state appears depending 
on the band filling\cite{Mic}. Various SC states can occur in the model with 
a single type of  (nearest-neighbor attractive) interaction because of 
the change of the shape of the Fermi surface\cite{KK3}.
Near half-filling ($\mu \sim 0$) the $d_{x^2-y^2}$-wave SC state is 
stabilized, while the extended $s$-wave SC state is favored near the band edge 
($\mu \sim \pm 4t$). In the region between $d$- and $s$-wave states 
spin-triplet $(p_x \pm ip_y)$-wave SC state appears. 

The properties of the interface states do not depend on the values of 
the parameters in a qualitative way, 
unless the symmetries of the superconducting ($d_{x^2-y^2}$ or 
$(p_x \pm ip_y)$-wave) and the magnetic (ferromagnetic or antiferromagnetic) 
states are changed. 
Therefore we will choose typical values of the parameters in order to realize 
the states in question. 
We will use finite values of $U_F$ to investigate the effect of 
electron correlations in the S side, while the attractive interaction in the 
F side is assumed to be absent, i.e., $V_F = 0$. 
Hereafter we take $t_F = t_S (\equiv t) =1$ as the unit of energy, 
and the tunneling matrix element $t_T$ is varied to see how the 
extent of the proximity effect is changed. 
\section{Interface States and Spontaneous Spin Currents for chiral p-wave 
superconductors} 

In this section we study the proximity effect between ferromagnets and  
superconductors with  $(p_x \pm ip_y)$-wave symmetry.    
Here the parameters are chosen to realize the ferromagnetic and 
$(p_x \pm ip_y)$-wave SC states in F and S layers, respectively, i.e., 
$U_F = 12$, $V_S = -2.5$ and $\mu = -1.6$. 
The system size we used is $N_x = N_y =120$, which is large enough 
to study the interface states, because the coherence lengths of SC and 
ferromagnetic states are of the order of $10a$ 
for the parameters used here. 
The spatial variations of the SCOP's and the magnetization 
$m$ near the [100] interface are shown in Fig.2 where 
the tunneling matrix element is taken to be $t_T = 1$, and $U_S = 1.5$.
It is seen that $\Delta_{px}$ and $\Delta_{py}$ are suppressed near 
the interface, and the spin-singlet components $\Delta_d$ and $\Delta_s$ 
are induced.  Near a surface of a chiral superconductor faced to vacuum, 
the spin-singlet components are not induced, 
though the $p$-wave SCOP's are suppressed also in this case\cite{matsu2}.
This difference is due to the presence of $m$ in the S side, 
because the SC state cannot be formed with $(p_x \pm ip_y)$-channel only, 
if $\langle n_\uparrow \rangle \not= \langle n_\downarrow \rangle$.
The proximity effects will be  reduced as the tunneling matrix element becomes 
smaller. This is actually the case as seen in Fig.3 where $t_T=0.1$ is used. 
It is seen that the penetration of $m$ into the S side is 
suppressed, and $d$- and $s$-wave SCOP's are much smaller compared 
with Fig.2. 
The change of the surface states, as $t_T$ is varied, is rather simple 
in the case of $(p_x \pm ip_y)$-wave superconductors. 
Namely, the state monotonously approaches that for $t_T = 0$ as $t_T$ 
is reduced. 
(On the contrary, the surface state of a $d$-wave superconductor shows a 
phase transition when $t_T$ is varied, as we will see in the next section.)
The penetration depth of $m$ depends on the electron correlation 
in the S side. The spatial variations of OP's for  $U_S=5$ and $t_T = 1$ 
are shown in Fig.4. 
For this value of $U_S$ the system gets closer to the magnetic instability 
 (though this $U_S$ is not large enough to stabilize the magnetic order), 
and the correlation length of $m$ becomes larger. 
Then the magnetization $m$ penetrates far inside the S side compared 
with Fig.2, and the $d$- and $s$-wave SCOP's decay rather slowly.

In the ${\cal T}$-violating SC state a spontaneous current flows along 
the surface in the region where the SCOP's are 
spatially varying\cite{Vol,SigR,SigUe}. 
Note that the condition for the minimum free energy requires 
the current normal to the interface to vanish. This condition in turn leads 
to a finite current along the interface if ${\cal T}$ is broken. 
(Namely the present spontaneous current is the equilibrium current.)
The expressions for the currents carried by spin-up and spin-down electrons 
are given by 
\begin{equation}\begin{array}{rl}
J_y^\uparrow(i) = & \displaystyle -\frac{2eta}{\hbar} \frac{1}{N_y} 
\sum_{k,n} \sin ka |u_{i,n}(k)|^2 f(E_n(k))   \\
J_y^\downarrow(i) = & \displaystyle \frac{2eta}{\hbar}\frac{1}{N_y} 
\sum_{k,n} \sin ka |v_{i,n}(k)|^2 [1-f(E_n(k))]. \\
\end{array}\end{equation}
Since we do not consider the effect of the vector potential ${\vec A}$ in 
this paper, terms linear in ${\vec A}$  in the expressions of $J_y$ 
are not taken into account. 
Near the surface faced to vacuum only the charge current 
(defined as $J_{\rm charge} = J^\uparrow + J^\downarrow$) can flow\cite{matsu2}.
When the surface is attached to a ferromagnet, 
$m$ is finite in the region where the chiral SCOP's have spatial variations. 
Thus $J^\uparrow$ is not equal to $J^\downarrow$, and 
the spontaneous spin current (defined as 
$J_{\rm spin} = J^\uparrow - J^\downarrow$) appears as seen in Fig.5(a).
For small value of $t_T (=0.1)$, only small spin current flows while 
the charge current is large (Fig5(b)). 
This is because the penetration of $m$ into the S side is 
suppressed (Fig.3), and then the difference between $J_\uparrow$ and 
$J_\downarrow$ is reduced. 
When the $(p_x \pm ip_y)$-wave superconductor is faced to vacuum, only 
a charge current arises and no spin current shows up \cite{matsu2}.
The $t_T$-dependence of spin currents presented here is consistent with 
this fact. 
When the effect of electron correlations is important in the S side, 
$J_{\rm charge}$ is reduced but $J_{\rm spin}$ is not (Fig.5(c)). 
This is because a large $U_S$ suppresses $\Delta_p$'s leading to a reduced 
$J_{\rm charge}$, while it enhances $m$ and $J_{\rm spin}$.  
The spin current also becomes larger as the value of $U_F$ is increased 
(Fig.5(d) with $U_F = 16$), since  $m$ in the F side is enhanced leading 
to larger value of $m$ in the S side. 
To summarize the discussion of the spin current in bilayer 
systems, $J_{\rm spin}$ can be large if the magnetization induced 
in the S side is large. 

We have also investigated the  case of [110] interface between 
a $(p_x \pm ip_y)$-wave superconductor and a ferromagnet, and the results 
are qualitatively the same as those for the [100] interface. 

Now we examine the F/S/F trilayer system where S is a chiral 
$(p_x \pm ip_y)$-wave superconductor. 
The spatial variations of the magnetization $m$ and the $p$-wave SCOP's 
are shown in Fig.6. (Small Re$\Delta_d$ and Re$\Delta_s$ are also 
induced near the interface as in F/S bilayer system, 
though they are not shown here.) 
In the F/S bilayer system, we have seen that spontaneous spin currents 
as well as charge currents flow along the interface. 
The direction of the charge current, $J_{\rm charge}$, is 
determined by the chirality (i.e., ($p_x + ip_y$) or $(p_x - ip_y$) ), 
and the currents at the opposite edges of the superconductor 
should flow in opposite directions. 
The direction of the spin current, $J_{\rm spin}$, is the same as (opposite to) 
that at the other edge of the superconductor if the magnetizations 
of two ferromagnets are antiparallel (parallel). 
Then if the chiral superconductor is sandwiched by ferromagnets with 
antiparallel magnetizations, the total charge current (integrated over $x$) 
should vanish, while the total spin current will be  finite. 
On the contrary if the magnetization of two F layers are parallel, 
both the total charge current and the total spin current should vanish.
This is actually the case as shown in Fig.7, where $\sum_i J_{\rm charge}(i)$ 
vanishes for both parallel and antiparallel configurations of $m$ and 
$\sum_i J_{\rm spin}(i)$ is finite only for an antiparallel configuration of $m$. 
If we assume $t = 1$eV, $\sum_i J_{\rm spin}(i) \sim 10 \mu$A.

\section{Interface States for d-wave superconductors} 
In this section we examine the interface states for ferromagnets and 
superconductors with $d_{x^2-y^2}$-wave symmetry. 
In order to realize the ferromagnetic and the $d_{x^2-y^2}$-wave SC states 
in F and S layers, respectively, the parameters are chosen to be $U_F = 10$, 
$U_S = 1.5$, $V_S = -1.5$ and $\mu = -0.5$. 
We consider only the [110] interface here, since a ${\cal T}$-breaking 
surface state may be formed in this case, but not in the case of 
a [100] surface\cite{Honer}.  
The ${\cal T}$-violating surface state can be formed by introducing additional 
SCOP, such as $is$ or $id_{xy}$-wave component\cite{Sig}.
In the present model $(d_{x^2-y^2}\pm is)$-wave state may occur, 
because the attractive interaction exists between nearest-neighbor sites 
on a square lattice.  
(It may be possible to induce $id_{xy}$ SCOP near the magnetic impurity 
with spin-orbit interactions or due to the magnetic field\cite{Balatsky,Laughlin}.)
In the $(d \pm is)$-wave surface state, the spontaneous charge current 
flows along the $y'$ direction.
If a ferromagnet is attached to a [110] surface of a $d_{x^2-y^2}$-wave 
superconductor,  the coexistence of the magnetization $m$ in the 
${\cal T}$-violating state may be expected, leading to a spontaneous spin current.

In Fig.8 the spatial variations of $m$ and the SCOP's near the [110] interface 
are shown for $t_T = 0.1$.  Since a $d_{x^2-y^2}$-wave SC state has nodes 
in contrast to a fully gapped $(p_x\pm ip_y)$-wave SC state, 
the proximity effect in the former case is much larger than that in the latter.
(See Fig.3 where the same $t_T = 0.1$ is employed in the case of 
$(p_x\pm ip_y)$-wave superconductor.)
When $t_T$ is not so small $(t_T \gtrsim 0.02$ for $U$'s and $V$'s used here), 
the magnetization penetrates into the S layer (Fig.9). 
When $m \not= 0$, the density of spin-up and spin-down electrons 
are not equal, and the Cooper pairs cannot be formed in singlet 
channels only, and thus the $p$-wave components have to arise\cite{KK1}.
We find that the complex component ($is$), which could be induced 
near a surface faced to vacuum, is destroyed, 
and the resulting state has a $(d_{x^2-y^2} + p_x + p_y)$-symmetry. 
When the value of $t_T$ is reduced,  the symmetry of the SC state changes.  
For $t_T \lesssim 0.02$, $(d \pm is)$-wave state appears, where 
the magnetization $m$  (and thus $p$-wave SCOP's) is absent in the S layer 
even though $t_T \not= 0$. (Fig.10) 

In the $(d \pm is)$-state a spontaneous charge current can flow along 
the interface, but there is no spin current because $m = 0$. 
In the $(d+p_x+p_y)$-state both charge and spin current vanish, 
because all SCOP's are real. (Fig.11)
For the [110] interface the expression for the current $J_{y'}^\sigma(i)$ is 
slightly different from that for [100] case. 
We checked that the current from the site $i$ to $i+{\hat x'}/2+{\hat y}'/2$, 
denoted as $J_a^\sigma(i)$, is the same as 
that from the site $i+{\hat x}'/2+{\hat y}'/2$ to $i+{\hat y}'$, 
denoted as $J_b^\sigma(i)$.   
(Otherwise $J_{x'}^\sigma(i) \not= 0$.) 
Then $J_{y'}^\sigma(i)$ is given by the $y'$ component of $J_a^\sigma(i)$ 
(or,  equivalently $J_b^\sigma(i)$): 
\begin{equation}\begin{array}{rl}
J_{y'}^\uparrow(i) = & \displaystyle \frac{1}{\sqrt 2}
\frac{2et_ia}{\hbar}\frac{1}{N_{y'}} \sum_{k,n} f(E_n(k)) \times \\ 
& \displaystyle Im [u^*_{i+\frac{a'}{2},n}(k)u_{i,n}(k) e^{-{\rm i}ka'/2}] \\
\\
J_{y'}^\downarrow(i) = & \displaystyle \frac{1}{\sqrt 2}
\frac{2et_ia}{\hbar}\frac{1}{N_{y'}} \sum_{k,n} [1-f(E_n(k))] \times \\ 
& \displaystyle Im [v_{i+\frac{a'}{2},n}(k)v^*_{i,n}(k)e^{{\rm i}ka'/2}]. 
\end{array}\end{equation}
where $t_i = t_T$ if $i$ corresponds to the interface site, $t_i = t$ otherwise.  

The transition between the $(d \pm is)$- and $(d+p_x+p_y)$-wave 
SC states is of first order, 
and we did not find a state which carries a spontaneous spin current 
for F/S($d_{x^2-y^2}$) interface states.  
Our results imply that both the state with 
$m \not= 0$, $\Delta_s = 0$ and $\Delta_{px(y)} \not= 0$, 
and the state with $m = 0$, $\Delta_s \not= 0$ and $\Delta_{px(y)} = 0$ 
correspond to the local minima of the free energy, and that the former (latter) 
has lower energy when $t_T > (t_T)_{cr}$ ($t_T < (t_T)_{cr}$) 
with $(t_T)_{cr} \sim 0.02$. 
Actually for $t_T \sim (t_T)_{cr}$, we obtained two kinds of solutions 
depending on the initial values of OP's for the iteration, and the energies of 
two solutions cross at $t_T = (t_T)_{cr}$. 

\section{Summary}

We have studied the proximity effect between unconventional 
superconductors and ferromagnets. It is found that a spontaneous spin current
can flow along the interface between a $(p_x \pm ip_y)$-wave superconductor 
and a ferromagnet due to the coexistence of the magnetization and the 
chiral superconducting state. 
In the case of a $d_{x^2-y^2}$-wave superconductor the induced magnetization 
destroys the ${\cal T}$-breaking $(d \pm is)$-wave surface state for relatively 
large transmission ($t_T \gtrsim 0.02$). For small transmission 
($t_T \lesssim 0.02$) the $(d \pm is)$-wave SC state with 
$m = 0$ is realized, and we did not find the state with a spontaneous 
spin current for any value of $t_T$. 
This implies that only a spontaneous charge current may be possible for 
high-$T_c$ cuprates, while both spin and charge currents may be expected 
for spin-triplet superconductors, e.g., Sr$_2$RuO$_4$.
For conventional $s$-wave superconductors 
based on the electron-phonon mechanism, the ferromagnetism plays 
a very strong role to suppress the superconductivity. 
On the contrary, in the case of unconventional superconductors 
these effects may be weaker, since the superconductivity 
is caused by the magnetic interactions.
Then we expect it is possible to obtain a surface state with a spontaneous spin 
current experimentally using spin-triplet superconductors. 

\section*{Acknowledgment}

The authors are grateful to G. Tatara, H. Kohno, M. Sigrist, H. Shiba and 
H. Fukuyama for useful discussions. This work was financially  
supported by Sumitomo foundation.

%%%%%%\newpage\centerline{\bf Figure Captions}

\medskip
\noindent 
{\bf Fig. 1}
Schematic descriptions of the interfaces between ferromagnets 
and superconductors for (a) [100] and (b) [110] surfaces. 

\medskip
\noindent 
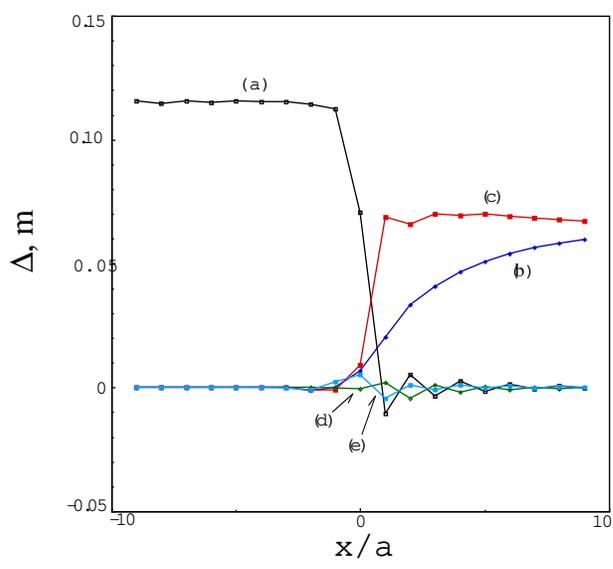
{\bf Fig. 2}  
Spatial variations of the magnetization $m$ and the SCOP's for a 
F/S$(p_x+ip_y)$ bilayer system with 
$t_T =1$, $U_F =12$, $U_S = 1.5$, 
$V_S = -2.5$, $\mu= -1.6$, $N_x = 60 + 60$ and $N_y =120$.  
(a) $m$, (b) Re$\Delta_{px}$, (c) Im$\Delta_{py}$, (d) Re$\Delta_d$ and 
(e) Re$\Delta_s$. Note all OP's are non-dimensional, and 
$x = 0$ corresponds to the interface site of a ferromagnet.  

\medskip
\noindent 
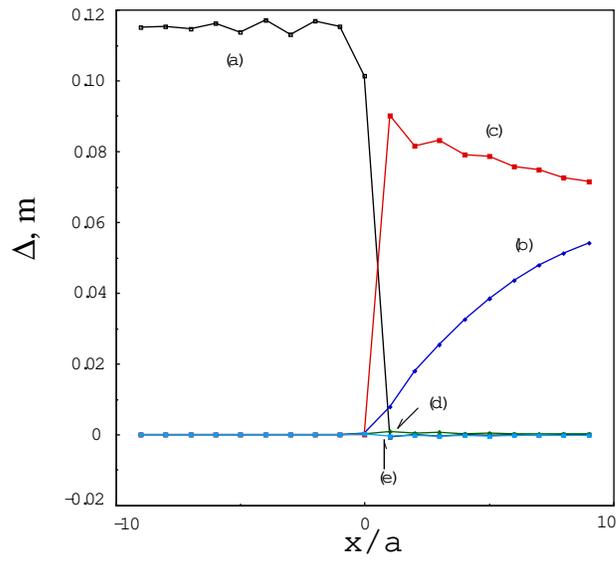
{\bf Fig. 3}  
Spatial variations of $m$ and the SCOP's for a 
F/S$(p_x+ip_y)$ bilayer system with $t_T=0.1$. 
Other parameters are the same as in Fig.2.
(a) $m$, (b) Re$\Delta_{px}$, (c) Im$\Delta_{py}$, (d) Re$\Delta_d$ 
and (e) Re$\Delta_s$. 

\medskip
\noindent 
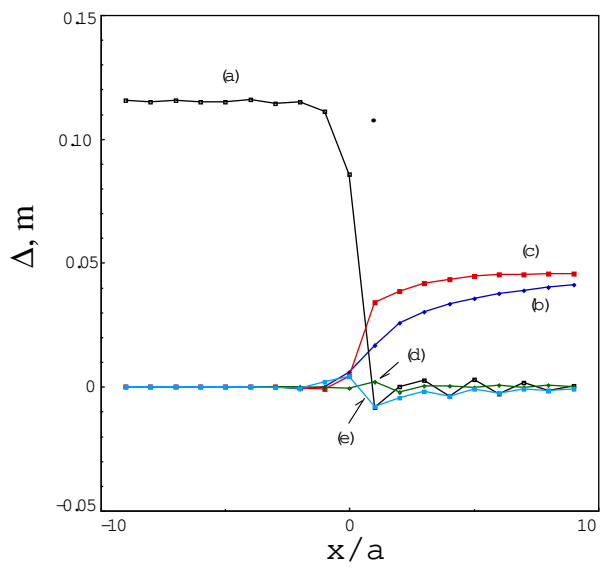
{\bf Fig. 4}  
Spatial variations of $m$ and the SCOP's for a 
F/S$(p_x+ip_y)$ bilayer system with $U_S=5$. 
Other parameters are the same as in Fig.2.
(a) $m$, (b) Re$\Delta_{px}$, (c) Im$\Delta_{py}$, (d) Re$\Delta_d$ 
and (e) Re$\Delta_s$. 

\medskip
\noindent 
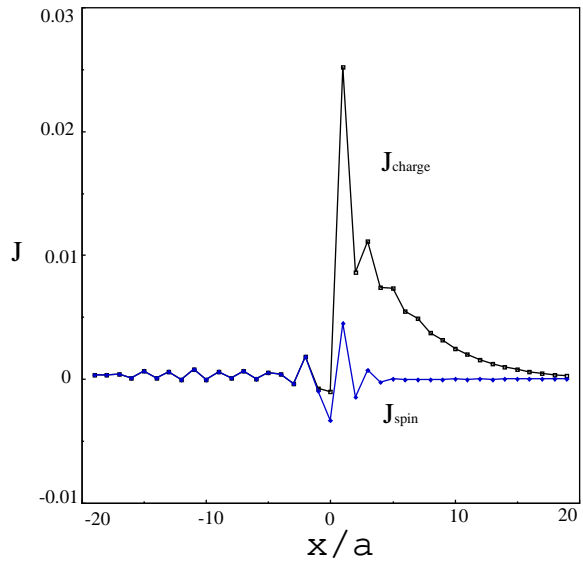
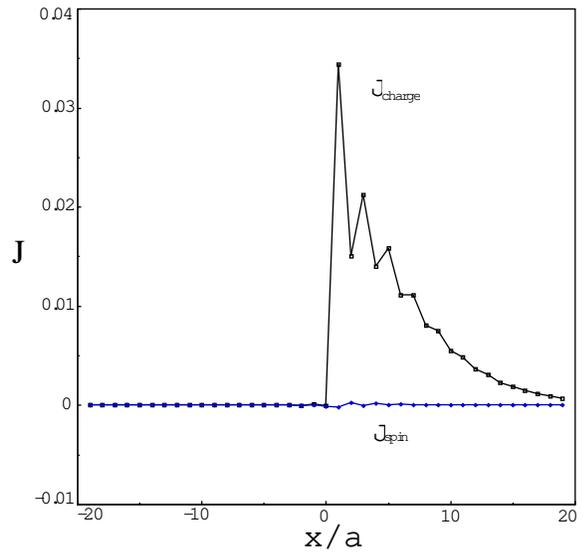
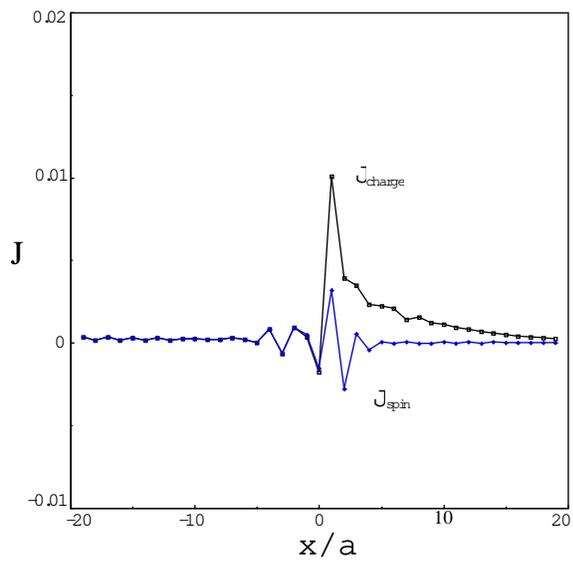
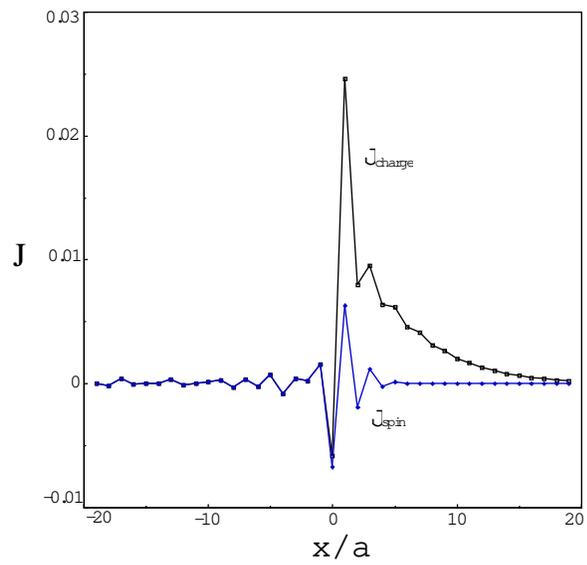
{\bf Fig. 5}  
The charge and the spin current (in units of $2eta/\hbar$) 
for F/S$(p_x+ip_y)$ bilayer systems. 
(a)  $U_F=12$, $U_S=1.5$ and $t_T=1$, (b)  $U_F=12$, $U_S=1.5$ and $t_T=0.1$, 
(c) $U_F=12$, $U_S = 5$ and $t_T = 1$ and (d) $U_F=16$, $U_S=1.5$ and $t_T=1$. 
Other parameters are the same as in Fig.2.

\medskip
\noindent 
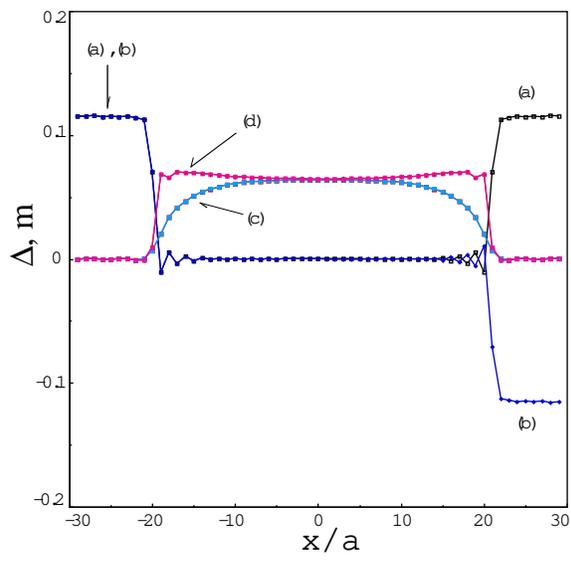
{\bf Fig. 6}  
Spatial variations of $m$, Re$\Delta_{px}$ and Im$\Delta_{py}$ 
for F/S$(p_x+ip_y)$/F trilayer systems. 
The system size is $N_x = 40 + 40 + 40$, $N_y =120$,  
and other parameters are the same as in Fig.2.
(a) $m$ with parallel configuration, (b) $m$ with antiparallel 
configuration in two F layers, respectively. 
(c) Re$\Delta_{px}$ and (d) Im$\Delta_{py}$.  
Note Re$\Delta_{px}$ and Im$\Delta_{py}$ do not depend on the 
configurations of $m$. 

\medskip
\noindent 
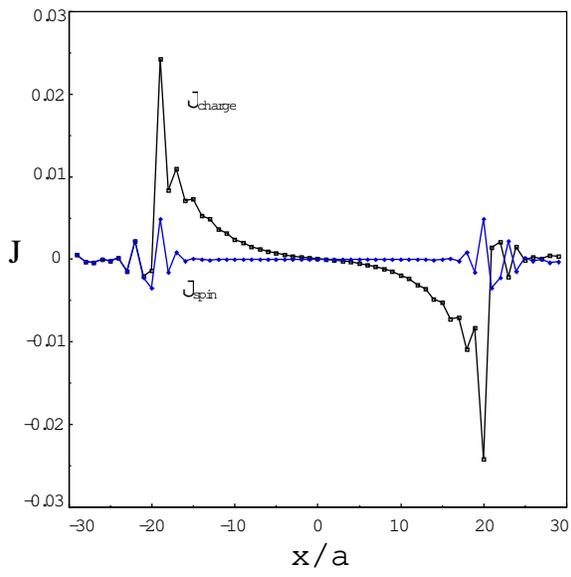
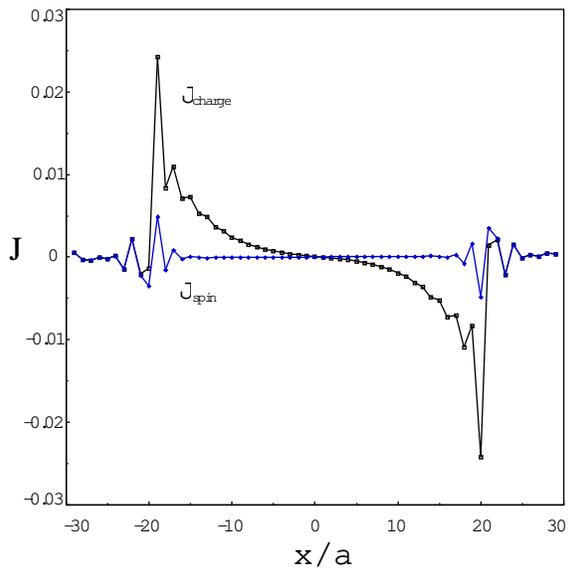
{\bf Fig. 7}   
The charge and the spin current  (in units of $2eta/\hbar$) 
for F/S$(p_x+ip_y)$/F trilayer systems. 
Parameters used are the same as in Fig.6.
Magnetization of two F layers are antiparallel in (a), 
and parallel in (b).

\medskip
\noindent 
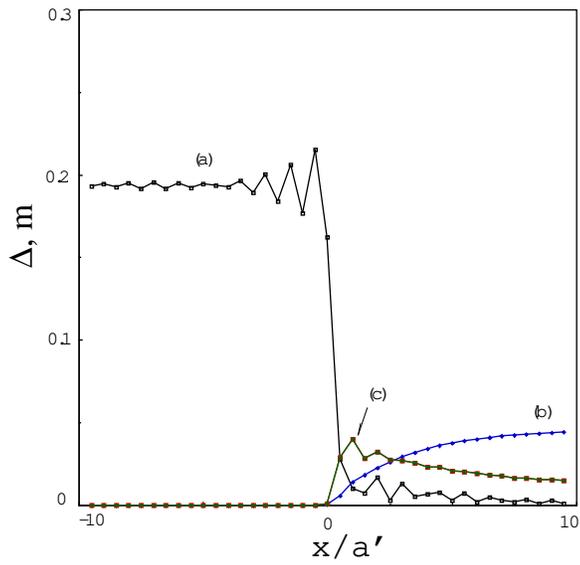
{\bf Fig. 8}
Spatial variations of $m$ and the SCOP's for a 
F/S($d_{x^2-y^2}$) bilayer system with 
$t_T = 0.1$, $U_F =10$, $U_S = 1.5$, 
$V_S = -1.5$, $\mu= -0.5$, $N_{x'} = 60 + 60$ and $N_{y'} = 120$. 
(a) $m$, (b) Re$\Delta_d$ and (c) Re$\Delta_{px} (=$ Re$\Delta_{py}$).

\medskip
\noindent 
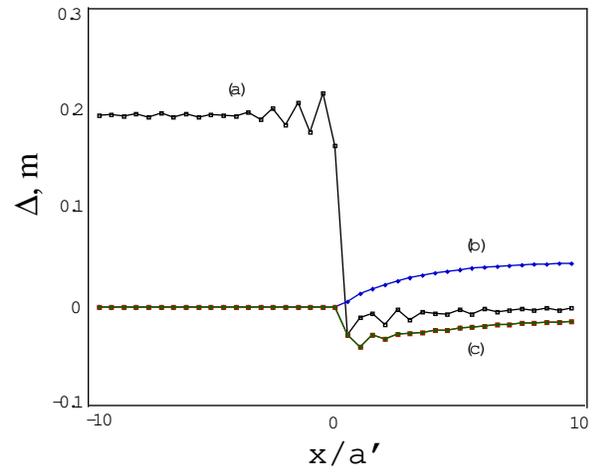
{\bf Fig. 9}
Spatial variations of $m$ and the SCOP's for a 
F/S($d_{x^2-y^2}$) bilayer system with $t_T=0.03$. 
Other parameters are the same as in Fig.8. 
(a) $m$, (b) Re$\Delta_d$ and (c) Re$\Delta_{px} (=$ Re$\Delta_{py}$).

\medskip
\noindent 
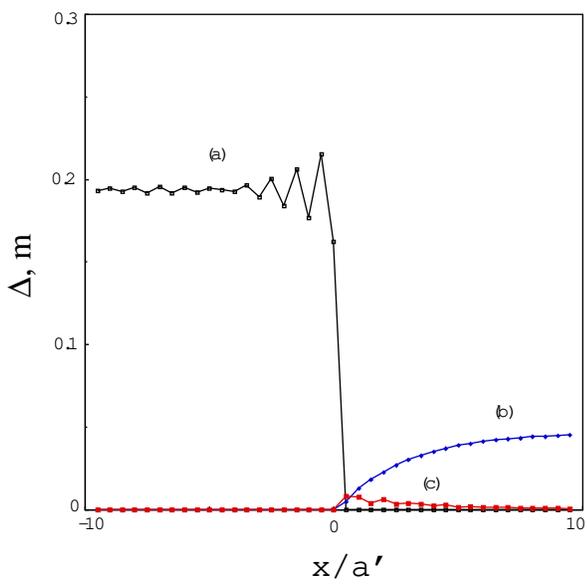
{\bf Fig. 10}
Spatial variations of $m$ and the SCOP's for a
F/S($d_{x^2-y^2}$) bilayer system with 
$t_T = 0.01$. Other parameters are the same as in Fig.8. 
(a) $m$, (b) Re$\Delta_d$ and (c) Im$\Delta_s$.

\medskip
\noindent 
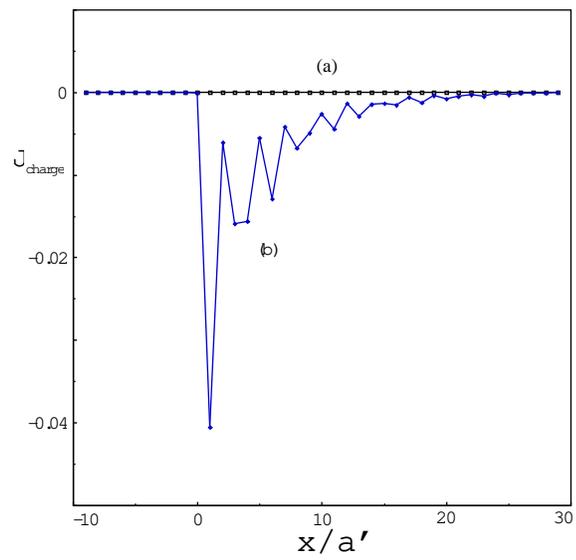
{\bf Fig. 11}
The charge current  (in units of $2eta/\hbar$) 
for F/S($d_{x^2-y^2}$) bilayer systems with (a) $t_T = 0.1$ and $t_T = 0.03$, 
and (b) $t_T = 0.01$.  
Other parameters are the same as in Fig.8.

%%%%%%%%%
\end{multicols}  
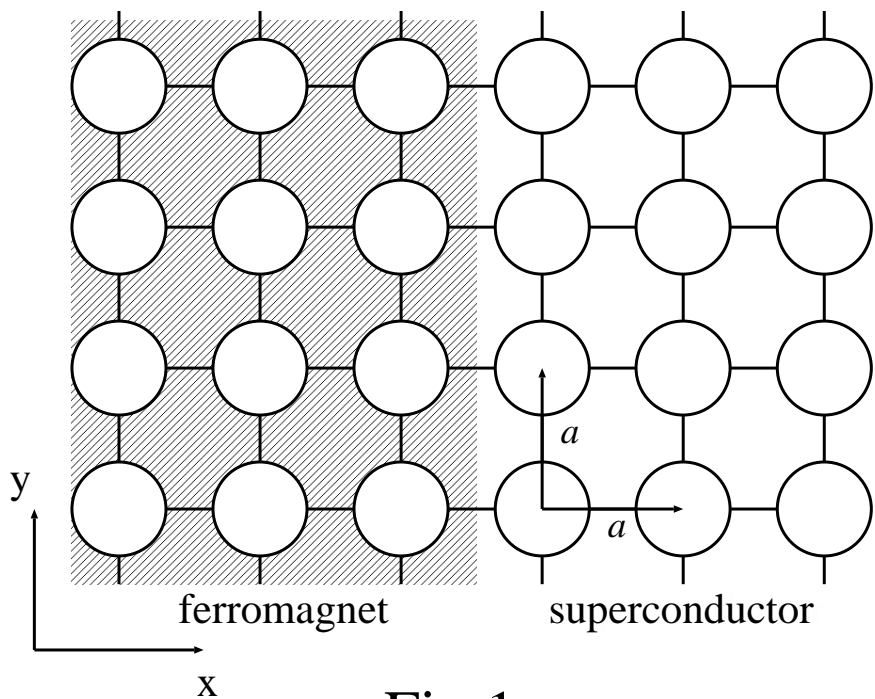
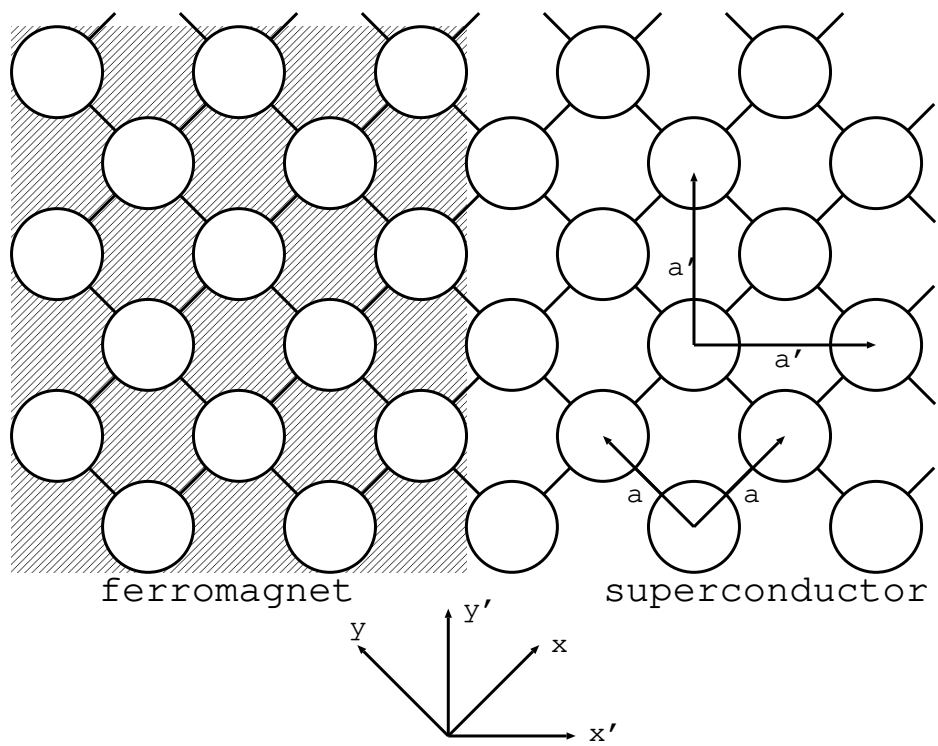
\end{document}